\documentclass[12pt,a4paper,onecolumn]{IEEEtran}
\usepackage[T1]{fontenc}
\usepackage[utf8]{inputenc}
\usepackage[english]{babel}
\usepackage{amsmath,amsfonts,amssymb,amsthm}
\usepackage{graphicx}
\usepackage{subcaption}
\usepackage{tikz}
\usepackage{cite}

\usetikzlibrary{shapes,arrows}

\newtheorem{Theorem}{Theorem}

\newtheorem{Assumption}[Theorem]{Assumption}

\begin{document}

\title{A framework for the development of intelligent mechanical systems}
\author{Wallace M.\ Bessa}
\date{}

\maketitle

\begin{abstract}
\noindent
From autonomous vacuum cleaners to self-driving cars, intelligent mechanical systems are becoming an intrinsic part of our daily lives. In this work, a framework for the development of intelligent mechanical systems is presented.Considering that in this scenario the adopted control approach plays an essential role, I show that the proposed scheme should be able to not only adapt itself to changes in the environment, but also learn from its own experience.
\end{abstract}

\section*{Introduction}

Intelligent control is usually defined as the discipline where control schemes are developed in order to emulate certain competencies of human intelligence~\cite{passino1995,antsaklis2001}.
However, even among psychologists and neuroscientists there is no consensus on what intelligence really means.
Some researchers suggest that prediction is the foundation of intelligence~\cite{llinas2001,Hawkins2004}; another widely assumed definition is the ability to adapt oneself adequately to new situations \cite{binet1916,pintner1921,garlick2002}; and it has also been characterized as the capacity to learn or to profit by experience~\cite{dearborn1921,hebb1949,Shanahan2015}.

Of course intelligence can have many other higher attributes, such as emotion~\cite{levine2009}, creativity~\cite{werbos2009} and consciousness~\cite{sanchez-canizares2014}, but the abilities to predict, adapt itself, and learn are the most basic ones, and can be found in all intelligent biological systems. 
They are also very convenient, from the control theory perspective, since they can provide a structured way to the development of intelligent schemes. 
On this basis, within the proposed framework, an intelligent controller should present at least the following three competencies:

\begin{description}[\IEEEsetlabelwidth{Adaptation}]
\item [Prediction] It should be able to incorporate some prior knowledge of the plant, in order to make plausible predictions about the expected dynamic behavior;
\item [Adaptation] It must adapt itself to changes in the plant and in the environment;
\item [Learning] It have to learn by experience as well as acquire knowledge by interacting with the environment, in order to improve the capacity to predict the dynamics of the plant. 
\end{description}

In this work, in order to comply with the ability to predict, a feedback linearization approach is used. 
Moreover, a single hidden layer neural network is embedded within the control law to fulfill with adaptation and learning.
The main details of the resulting scheme is presented in the next section.

\section*{Intelligent feedback linearization}

Intelligent control has proven to be a very attractive approach to cope with uncertain nonlinear systems
\cite{Bess2005,Bessa2005b,Bessa2010,Bessa2012,Tanaka2013,Bessa2014,Bessa2017,Bessa2018,Lima2018,Bessa2019,Deodato2019,Lima2020,Lima2021}. 
By combining nonlinear control techniques, such as feedback linearization or sliding modes, with adaptive intelligent algorithms, for example fuzzy logic or artificial neural networks, the resulting intelligent control strategies can deal with the nonlinear characteristics as well as with modeling imprecisions and external disturbances that can arise.

Now, Consider a class of $n^\mathrm{th}$-order nonlinear systems:

\begin{equation}
x^{(n)}=f(\mathbf{x},t)+b(\mathbf{x},t)u+d
\label{eq:system}
\end{equation}

\noindent
where $u$ is the control input, the scalar variable $x$ is the output of interest, $x^{(n)}$ is the $n$-th time 
derivative of $x$, $\mathbf{x}=[x,\dot{x},\ldots,x^{(n-1)}]$ is the system state vector, $f,b:\mathbb{R}^n\to
\mathbb{R}$ are both nonlinear functions and $d$ is assumed to represent all uncertainties and unmodeled dynamics 
regarding system dynamics, as well as any external disturbance that can arise. 

In respect of the disturbance-like term $d$, the following assumption will be made:

\begin{Assumption}
The disturbance $d$ is unknown but continuous and bounded, \textit{i.\,e}.\ $|d|\le\delta$.
\label{as:limd}
\end{Assumption}

Let me now define an appropriate control law based on conventional feedback linearization scheme that ensures the 
tracking of a desired trajectory $\mathbf{x}_d=[x_d,\dot{x}_d,\ldots,x^{(n-1)}_d]$, \textit{i.\,e}.\ the controller 
should assure that $\mathbf{\tilde{x}}\rightarrow0$ as $t\rightarrow\infty$, where $\mathbf{\tilde{x}}=\mathbf{x}
-\mathbf{x}_d=[\tilde{x},\dot{\tilde{x}},\ldots,\tilde{x}^{(n-1)}]$ is the related tracking error. 

On this basis, assuming that the state vector $\mathbf{x}$ is available to be measured and system dynamics is 
perfectly known, \textit{i.\,e}.\ there is no modeling imprecision nor external disturbance ($d=0$) and the functions 
$f$ and $b$ are well known, with $|b(\mathbf{x},t)|>0$, the following control law:

\begin{equation}
u=b^{-1}(-f+x^{(n)}_d-k_0\tilde{x}-k_1\dot{\tilde{x}}-\cdots-k_{n-1}\tilde{x}^{(n-1)})
\label{eq:lawfl}
\end{equation}

\noindent 
guarantees that $\mathbf{x}\rightarrow\mathbf{x}_d$ as $t\rightarrow\infty$, if the coefficients $k_i$ $(i=0,2,
\ldots,n-1)$ make the polynomial $p^n+k_{n-1}p^{n-1}+\cdots+k_0$ a Hurwitz polynomial \cite{Slotine1991}.

The convergence of the closed-loop system could be easily established by substituting the control law (\ref{eq:lawfl}) 
in the nonlinear system (\ref{eq:system}). The resulting dynamical system could be rewritten by means of the tracking 
error:

\begin{equation}
\tilde{x}^{(n)}+k_{n-1}\tilde{x}^{(n-1)}+\ldots+k_1\dot{\tilde{x}}+k_0\tilde{x}=0
\label{eq:cl}
\end{equation}

\noindent 
where the related characteristic polynomial is Hurwitz.

However, since in real-world applications the nonlinear system (\ref{eq:system}) is often not perfectly known, the 
control law (\ref{eq:lawfl}) based on conventional feedback linearization is not sufficient to ensure the exponential 
convergence of the tracking error to zero. 

\begin{figure}[htb]
\centering
\begin{tikzpicture}
[   cnode/.style={draw=black,fill=blue!20,minimum width=#1,circle},
    rnode/.style={draw=black,fill=blue!20,minimum width=#1,rectangle},
]
    \node[rnode=2mm,label=180:$s$] (i) at (0,-3) {};
    \node[cnode=6mm] (f1) at (2,-1) {$\varphi_1$};
    \node[cnode=6mm] (f2) at (2,-2) {$\varphi_2$};
    \node[cnode=6mm] (f3) at (2,-3) {$\varphi_3$};
    \node[cnode=6mm] (fn) at (2,-5) {$\varphi_n$};
    \node[cnode=6mm] (s) at (5,-3) {$\Sigma$};
    \node[right] at (6,-3) {$\hat{d}$};
    \node at (2,-3.9) {$\vdots$};
    \draw [-latex'] (i) -- (f1);
    \draw [-latex'] (i) -- (f2);
    \draw [-latex'] (i) -- (f3);
    \draw [-latex'] (i) -- (fn);
    \draw [-latex'] (f1) -- node[above,sloped,pos=0.4] {$w_1$} (s);
    \draw [-latex'] (f2) -- node[above,sloped,pos=0.4] {$w_2$} (s);
    \draw [-latex'] (f3) -- node[above,sloped,pos=0.4] {$w_3$} (s);
    \draw [-latex'] (fn) -- node[above,sloped,pos=0.4] {$w_n$} (s);
    \draw [-latex'] (s) -- (6,-3); 
    \node at (0,-0.2) {Input};
    \node at (2.2,-0.2) {Hidden layer};
    \node at (5.2,-0.2) {Output};
\end{tikzpicture}
\caption{Single hidden layer feedforward neural network.}
\label{fig:shln}
\end{figure}
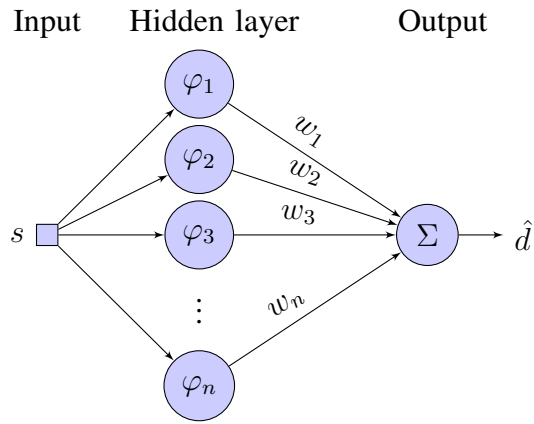

Considering that neural networks can perform universal approximation \cite{Hornik1991}, I propose the adoption of a single 
hidden layer feedforward neural network within the control law, in order to compensate for $d$ and to enhance the feedback 
linearization controller. The chosen network architecture is presented in Fig.~\ref{fig:shln}, with Gaussian functions as 
neural basis functions, $\varphi_i$, and a filtered tracking error, $s$, as input.

Therefore, the control law with the fuzzy compensation scheme can be stated as follows

\begin{equation}
u=b^{-1}[-f+x^{(n)}_d-k_0\tilde{x}-k_1\dot{\tilde{x}}-\cdots-k_{n-1}\tilde{x}^{(n-1)}-\hat{d}(\mathbf{\tilde{x}})]
\label{eq:lawfuzzy}
\end{equation}

\noindent
and the related closed-loop system is:

\begin{equation}
\tilde{x}^{(n)}+k_{n-1}\tilde{x}^{(n-1)}+\ldots+k_1\dot{\tilde{x}}+k_0\tilde{x}=\tilde{d}
\label{eq:clwd}
\end{equation}

\noindent
with $\tilde{d}=\hat{d}-d$.

Now, defining $\mathbf{k^\mathrm{T}\tilde{x}}=k_{n-1}\tilde{x}^{(n-1)}+\ldots+k_1\dot{\tilde{x}}+k_0\tilde{x}$,
where $\mathbf{k}=[c_0\lambda^n, c_1\lambda^{n-1},\dots,c_{n-1}\lambda]$, $\lambda$ is a strictly positive
constant and $c_i$ states for binomial coefficients, \textit{i.\,e}.\

\begin{equation}
c_i=\binom{n}{i}=\frac{n!}{(n-i)!\:i!}\:,\quad i=0,1,\ldots,n-1 
\label{eq:binom}
\end{equation}

\noindent 
the convergence of the closed-loop signals to a bounded region is assured. 

More details about the boundedness and convergence properties of the resulting controller, as well as some numerical 
simulations to illustrate the overall system performance, should be presented in the final version of the manuscript.

\section*{Acknowledgments}

This work was supported by the Alexander von Humboldt Foundation and the Brazilian Coordination for the Improvement of Higher Education Personnel.

\end{document}